\documentstyle[pra,aps]{revtex}
\begin{document}
\draft
\twocolumn
\title{Quantum jumps in hydrogen-like systems}
\author{Thorsten K\"ohler}
\address{ Institut f\"ur Theoretische Physik,
  Universit\"at G\"ottingen, Bunsenstr. 9,
  D-37073 G\"ottingen, Germany}
\maketitle
\begin{abstract}
In this paper it is shown that the Lyman-$\alpha$
transition of a single hydrogen-like system 
driven by a laser exhibits macroscopic dark periods,
provided there exists an additional constant 
electric field.
We describe the photon-counting process  
under the condition that the polarization of
the laser coincides with the direction of the constant
electric field.
The theoretical results are given for the example
of $^4\mathrm{He}^+$.
We show that the emission behavior depends sensitively
on the Lamb shift (W.~E.~Lamb, R.~C.~Retherford, Phys.~Rev.
  $\mathbf{72}$, 241 (1947)) 
between the $2s_{\frac{1}{2}}$ and $2p_{\frac{1}{2}}$ energy levels.
A possibly realizable measurement 
of the mean duration of the dark periods
should give quantitative information about the above 
energy difference by using the proposed photon-counting process.
\end{abstract}
\section{Introduction}
For the first time coherence effects in hydrogen-like systems
were found by the observation of quantum beats (Stark beats)
in the Lyman-$\alpha$ transition
\cite{Andrae: Stark Beats Artikel}.
In these experiments only the metastable $2s$ state of
atomic hydrogen is initially populated.  
Switching on a constant electric field
leads to a build-up of a coherent superposition 
of the upper $2p$ and $2s$ levels,
and the radiative decay shows an interference pattern
known as quantum beats.

This uncommon behavior of hydrogen-like systems suggests
that interesting effects may occur if the 
Lyman-$\alpha$ transition is driven by monochromatic 
laser light and if an additional constant electric field 
leads to a coherence between the upper levels
$2p$ and $2s$.
We show that the resulting photon-counting process
is similar to the one predicted by Dehmelt 
\cite{Dehmelt Quantenspruenge}
for a different system with two excited states,
one rapidly decaying and the other metastable,
driven by two lasers. Semiclassically one expects 
for the Dehmelt system periods of constant
fluorescence intensity due to the strong transition
(light period), interrupted by periods of zero intensity,
while the atomic electron is shelved in the 
metastable state (dark period).
These photon statistics have been observed experimentally
\cite{Dehmelt System Experimente},
and the above semiclassical idea has been analyzed
quantum mechanically
\cite{QM Beschreibung der Leucht und Dunkelphasen,%
Quantensprungmethode,Emissionsprozess: Cohen Tannoudji}.
In an alternative experiment Hulet and Wineland proved
the existence of macroscopic dark periods in the 
fluorescence intensity of a single ion
influenced by a magnetic field,
when a single laser is tuned near one of the 
principal transition resonances
\cite{alternatives Experiment von
  Hulet und Wineland}.
Hegerfeldt and Plenio 
\cite{Hegerfeld Plenio:
  Dunkelphasen ohne metastabiles Niveau}
proposed another mechanism
for dark periods which is not based on the existence of 
a metastable state.
They studied a three-level atom with two strong   
electric dipole transitions to one common ground state,
driven by a single laser. The existence of macroscopic
dark periods in the fluorescence light is due to a 
quantum coherence effect. The premise is a very small
energy separation of the upper levels in conjunction
with parallel transition dipole moments.
Because of this an experimental realization of the latter 
physical system seems to be difficult
\cite{Hegerfeldt Plenio:
  Dunkelphasen in
  Lambda-Systemen}.

In this paper we discuss the photon-counting process
of a single hydrogen-like system, driven by a single linearly polarized
laser and additionally influenced by a constant electric field.
As for the Dehmelt system, there exists a semi classical 
explanation for the occurrence of macroscopic dark periods
in the emission process of the hydrogen-like system
as follows.
The strong Lyman-$\alpha$ transition ($2p\rightarrow 1s$)
is driven by the laser light. Because of this we expect
a constant fluorescence intensity.
The constant electric field leads to the possibility
that the atomic electron makes a transition from the $2p$ to the 
$2s$ energy level.
In this case we have zero intensity (dark period),
because a dipole transition to the $1s$ ground state
is impossible.
But on the other hand, due to the constant electric field, there 
exists the possibility that the atomic electron gets out of
the $2s$ back into the $2p$ energy level, and the emission process
starts again.
By the quantum mechanical treatment of the problem we show
that under the assumption that the polarization of the laser 
coincides with the direction of the constant electric field
the above semiclassical explanation describes the photon-emission
process qualitatively
\cite{Cook Kimble: Telegraphenprozess}.

As will be seen later, one can regulate the mean duration of the 
dark and light periods almost independently by varying the intensity 
of the laser beam and the strength of the constant electric field.
The mean duration of the dark periods depends sensitively on
the Lamb shift
\cite{Lamb Retherford} 
between the $2s_{\frac{1}{2}}$ and $2p_{\frac{1}{2}}$ energy level.
A possibly realizable measurement of this mean duration, 
as it was done in the experiments
\cite{Dehmelt System Experimente,alternatives Experiment von
  Hulet und Wineland} in the case of other systems,
should give quantitative information about the above energy 
difference.

We show that there exists a correspondence between this
system and the above mentioned mechanism of macroscopic 
dark periods without a metastable state 
\cite{Hegerfeld Plenio:
  Dunkelphasen ohne metastabiles Niveau}.

\section{Quantum mechanical description of the 
photon-counting process}
We consider a single hydrogen-like system 
without hyperfine structure 
\cite{Hyperfeinstruktur}
driven by a single linearly polarized laser 
with electric field ${\bf F}_L$ and 
additionally influenced by a weak constant 
electric field ${\bf F}$.
The laser is supposed to be tuned near the 
$2p_{\frac{1}{2}}\rightarrow 1s_{\frac{1}{2}}$
transition resonance.
The Hamiltonian in dipole form for the atom 
interacting with the quantized radiation field
is given by 
\begin{eqnarray}
  \nonumber
  H=H_{A} &+& \sum_{{\bf k}\lambda}\hbar\omega_k
  a_{{\bf k}\lambda}^{\dagger}
  a_{{\bf k}\lambda}\\
  \nonumber
  &+& \sum_{{\bf k}\lambda}{\rm i}\sqrt{\frac{\hbar\omega_k}
    {2\varepsilon_0 V}}
  \left(
    a_{{\bf k}\lambda}
    -a_{{\bf k}\lambda}^{\dagger}
  \right)
  e{\bf D}\cdot
  {\mbox{\boldmath$\varepsilon$}}_{\mathbf{k}\lambda}\\
  \label{Gesamthamiltonoperator}
  &+& e{\bf D}\cdot
  \left(
    {\bf F}_L(t)
    +{\bf F}
  \right).
\end{eqnarray}
Here $H_A$ is the atomic fine structure Hamiltonian
\cite{Bethe Salpeter}.
We assume the Lamb shift to be incorporated in
$H_A$
\cite{Loudon Lehrbuch}.
As in Fig.\ \ref{fig1} the relevant atomic eigenstates with positive
magnetic quantum number are labeled from
$|1\rangle$ to $|5\rangle$.
For every $|i\rangle$ ($i=1,\ldots ,5$) with positive 
magnetic quantum number $m_i$,
there exists a corresponding atomic eigenstate with the same 
principal quantum number, the same total angular momentum
quantum number, the same parity
and the magnetic quantum number $-m_i$,
which we denote by $|-i\rangle$.
Then the atomic Hamiltonian is given by
\begin{equation}
  H_A=\sum_{|i|=1}^5 \hbar\omega_{i1} |i\rangle\langle i|,
\end{equation}
where $\omega_{ij}$ is the transition frequency between the 
states $|i\rangle$ and $|j\rangle$.
Note that $|i|>|j|$ implies $\omega_{i1}\geq\omega_{j1}$.

To describe the photon-counting process one
needs the probability density
$w(t_1,\ldots ,t_N;\left[t_0 ,t\right])$
for finding exactly $N$ photons at times
$t_1<\cdots <t_N$ in the interval $\left[t_0 ,t\right]$.
We assume that the initial state of the complete system is
$|\mathit{\Omega}\rangle\langle\mathit{\Omega}|\otimes\rho(t_0)$,
where $|\mathit{\Omega}\rangle$ is the vacuum of 
the quantized radiation 
field, and $\rho(t_0)$ is the initial atomic density operator.
The probability density $w$ is of the form
\cite{Davies Srinivas,Gardiner: Quantum noise,Hegerfeldt reset}
\begin{eqnarray}
  \nonumber
  w(t_1,\ldots ,t_N;\left[t_0,t\right])
  ={\rm Tr}\left(\hat{S}(t,t_N)\right.&\hat{J}&\hat{S}(t_N,t_{N-1})
  \times
  \\
  \label{gesuchte Wahrscheinlichkeitsdichte}
  \times
  \cdots&\hat{J}&
  \left.\hat{S}(t_1,t_0)\rho(t_0)\right),
\end{eqnarray}
where $\hat{S}(t,t^\prime)$ and $\hat{J}$ are atomic superoperators.
Using the quantum jump approach
\cite{Quantensprungmethode}
(which is essentially equivalent to 
the Monte Carlo wave function approach
\cite{MCWA} and to the use of quantum trajectories
\cite{Carmichael Quantum Trajectories})
Hegerfeldt 
\cite{Hegerfeldt reset}
has determined the superoperators $\hat{J}$
and $\hat{S}(t,t^\prime)$ for an arbitrary atom.
Considering Refs.
\cite{Quantensprungmethode,Hegerfeldt reset}
one finds that 
$|\mathit{\Omega}\rangle\langle\mathit{\Omega}|\otimes
\left(
  \hat{J}\rho(t)/{\rm Tr}(\cdot)
\right)$ 
is the state right after the detection of a photon,
provided we have a measurement by absorption, which we 
assume from now on.
There exists a nonunitary atomic operator 
$U_c(t,t_0)$, the so called conditional (reduced)
evolution operator, describing the time development
of an atom under the condition that no photon is observed
in the time interval $[t_0,t]$,
and $\hat{S}(t,t_0)$ is given by
$\hat{S}(t,t_0)\rho(t_0)=
U_c(t,t_0)\rho(t_0)U_c(t,t_0)^\dagger$.
The conditional evolution operator is generated by 
the conditional Hamiltonian $H_c$ and 
with ${\bf D}_{ij}\equiv\langle i|{\bf D}|j\rangle$,
the generalized damping terms 
\begin{equation}
  \Gamma_{ijkl}\equiv
  \frac{e^2}{6\pi\varepsilon_0\hbar c^3}
  {\bf D}_{ij}\cdot {\bf D}_{kl} |\omega_{kl}|^3
\end{equation}
and the decay matrix 
\begin{equation}
  \Gamma\equiv\sum_{|i|,|j|,|k|=1
    \atop{|i|,|k|>|j|}}^5
  \Gamma_{ijjk} |i\rangle\langle k|
\end{equation}
one finds following
Refs.
\cite{Quantensprungmethode,Hegerfeldt reset}
that $H_c$ is given by
\begin{equation}
  \label{bedingter Hamiltonian}
  H_c=H_A+e{\bf D}\cdot
  \left(
    {\bf F}_L(t)+{\bf F}
  \right)
  -{\rm i}\hbar\Gamma.
\end{equation}
For the reset operator $\hat{J}$ one obtains
\cite{Hegerfeldt reset}
\begin{equation}
  \hat{J}\rho=\sum_{ijkl\atop{|i|>|j|\atop{|k|>|l|}}}
  \left(
    \Gamma_{jikl}
    +\Gamma_{klji}
  \right)
  \langle i|\rho| k\rangle
  |j\rangle\langle l|.
\end{equation}
The above operators $H_c$ and $\hat{J}$ are rather
complicated in the case of our ten level 
hydrogen-like system.
There are three main difficulties in calculating the probability
density Eq.
(\ref{gesuchte Wahrscheinlichkeitsdichte}) as follows.
The superoperator $\hat{J}$ generally carries pure states
into statistical mixtures of the two ground states
$|1\rangle$, $|-1\rangle$, with in general non zero off-diagonal
matrix elements.
The state right after the detection of a photon 
$|\mathit{\Omega}\rangle\langle\mathit{\Omega}|\otimes
\left(
  \hat{J}\rho(t)/{\rm Tr}(\cdot)
\right)$ 
generally depends on $\rho(t)$ 
\cite{Erklaerung zum Ruecksetzoperator}.
The superoperators $\hat{J}$ and $\hat{S}(t,t^\prime)$
act on a 100 dimensional vector space.
In the next section we show a way out of these difficulties.

\section{Symmetry considerations}
Here we make the assumption that the laser light
is linearly polarized with polarization in the same 
direction as the constant electric field,
${\bf F}=F{\bf e}_z$ and
${\bf F}_L(t)=F_L \cos(\omega_L t+\varphi_L)
{\bf e}_z$.
Therefore we have an invariance of the Hamiltonian
Eq. (\ref{Gesamthamiltonoperator})
with respect to the group $O(2)$ of those orthogonal
transformations which leave the $z$-axis invariant.
The results of the following symmetry considerations
are given at the end of this section.

By $T$ we denote the standard double-valued representation
of the group $O(2)$ on the atomic Hilbert space.
The invariance of the Hamiltonian
Eq. (\ref{Gesamthamiltonoperator})
leads to 
\begin{equation}
  \label{Invarianzeigenschaft von J und S}
  T(g)\left(\hat{J}\rho\right)T(g)^\dagger
  =\hat{J}\left(T(g)\rho T(g)^\dagger\right)
\end{equation}
for every group element $g$ and to the analogous 
equation for $\hat{S}(t,t^\prime)$.
Special elements of $O(2)$ are given by ${\cal D}(\varphi)$,
the rotation through the angle $\varphi$ about the $z$-axis,
and by ${\cal S}$, the reflection in the $y$-$z$ plain.
One can show that the projector $\hat{P}^{(0)}$
onto the subspace of the scalar operators 
with respect to the group $O(2)$ is given by
\cite{Darstellungstheorie} 
\begin{eqnarray}
  \nonumber
  &&\hat{P}^{(0)}\rho=\frac{1}{4\pi}
  \int_0^{2\pi} 
  {\rm d}\varphi
  \left(
    T({\cal D}(\varphi))\rho T({\cal D}(\varphi))^\dagger
  \right.\\
  &&\hspace{35pt}
  \left.
    \nonumber
    +T({\cal D}(\varphi))T({\cal S})\rho 
    T({\cal S})^\dagger T({\cal D}(\varphi))^\dagger
  \right)\\
  \nonumber
  &&=\frac{1}{2}
  \sum_{{i,j>0}
    \atop{m_i=m_j}}
  \left(
    \langle i|\rho|j\rangle+\varepsilon_i^*\varepsilon_j
    \langle -i|\rho|-j\rangle
  \right)\\
  \label{Projektor auf Skalare}
  &&\hspace{49pt}\times
  \left(
    |i\rangle\langle j|
    +\varepsilon_i\varepsilon_j^*
    |-i\rangle\langle -j|
  \right),
\end{eqnarray}
where $\rho$ is an arbitrary atomic operator and
$\varepsilon_i$ is a phase factor with the 
property $T({\cal S})|i\rangle=\varepsilon_i |-i\rangle$.
Using the cyclic invariance of the trace and
Eq. (\ref{gesuchte Wahrscheinlichkeitsdichte}),
Eq. (\ref{Invarianzeigenschaft von J und S}) 
we obtain
\begin{eqnarray}
  \nonumber
  w(t_1,\ldots,t_N;[t_0,t])={\rm Tr}
  \left(
    \hat{S}(t,t_N)\hat{J}\hat{S}(t_N,t_{N-1})
  \right.\\
  \label{erste Vereinfachung von w}
  \left.
    \times\cdots\hat{J}\hat{S}(t_1,t_0)\hat{P}^{(0)}\rho(t_0)
  \right).
\end{eqnarray}
Since the atomic state right after the detection of a 
photon is a statistical mixture of the ground states 
$|1\rangle$, $|-1\rangle$
one finds
\begin{equation}
  \label{vereinfachter Ruecksetzzustand}
  \hat{J}\hat{P}^{(0)}\rho
  ={\rm Tr}\left(\hat{J}\rho\right)\rho_r
\end{equation}
with the scalar operator
\begin{equation}
  \rho_r=\frac{1}{2}
  (|1\rangle\langle 1|+|-1\rangle\langle -1|).
\end{equation}
From Eq. (\ref{Invarianzeigenschaft von J und S})
we also know that $\hat{J}$ and $\hat{S}(t,t^\prime)$
leave the subspace of the scalar operators 
invariant, and using additionally Eq.
(\ref{vereinfachter Ruecksetzzustand})
we come to the conclusion that the probability density
Eq. (\ref{gesuchte Wahrscheinlichkeitsdichte})
factorizes into single-photon probabilities
\begin{eqnarray}
  \nonumber
  &&w(t_1,\ldots ,t_N;\left[t_0,t\right])
  ={\rm Tr}\left(\hat{S}(t,t_N)\rho_r\right)\\
  \nonumber
  &&\times
  {\rm Tr}
  \left(\hat{J}\hat{S}
    (t_N,t_{N-1})\rho_r
  \right)
  \cdots
  {\rm Tr}
  \left(
    \hat{J}\hat{S}(t_2,t_1)\rho_r
  \right)\\
  \label{vereinfachte W-Dichte}  
  &&\times
  {\rm Tr}\left(\hat{J}\hat{S}(t_1,t_0)\hat{P}^{(0)}\rho(t_0)\right).
\end{eqnarray}
From this we see that the photon-counting process is 
governed by $P_0(t)$, the probability density of counting
no photons until $t$ starting with the ground state, since 
$-\dot{P}_0(t-t^\prime)={\rm Tr}
\left(\hat{J}\hat{S}(t,t^\prime)\rho_r\right)$
\cite{Hegerfeldt reset}.
The conditional evolution operator $U_c(t,t^\prime)$ 
is a scalar operator and from Eq. (\ref{Projektor auf Skalare})
one obtains that $U_c(t,t^\prime)$ leaves the
atomic subspace generated by the states with
positive magnetic quantum number invariant.
Using the definition 
of $\rho_r$, the relation
$T({\cal S})|1\rangle\langle 1|T({\cal S})^\dagger=
|-1\rangle\langle-1|$,
the fact that $U_c(t,t^\prime)$ commutes with
$T({\cal S})$ and the cyclic invariance of the trace
we obtain
\begin{eqnarray}
  \nonumber
  P_0(t)&=&{\rm Tr}\left(\hat{S}(t,0)\rho_r\right)\\
  \nonumber
  &=&{\rm Tr}\left(U_c(t,0)
  \rho_r U_c(t,0)^\dagger\right)\\
  &=&\| U_c(t,0)|1\rangle\|^2.
\end{eqnarray}
Therefore we only have to consider the atomic states with
positive magnetic quantum number. By a similar reasoning 
it is easy to check that the latter is also true in the more 
general case of the first single photon probability
${\rm Tr}(\hat{J}\hat{S}(t_1,t_0)\hat{P}^{(0)}\rho(t_0))$ in Eq.
(\ref{vereinfachte W-Dichte}). 

Because of our symmetry assumption 
the original ten-level atom behaves like
a five-level system with a single ground state
\cite{Verallgemeinerung}.
In the special case of the spontaneous Lyman-$\alpha$
transition influenced by the Stark effect this 
agrees with the previous results in literature 
(see for example 
\cite{Lueders Lebensdauer}).
Starting with the state right after the detection of a
photon even the state $|5\rangle$ drops out, because
it is coupled neither by the laser nor by the constant
electric field.
Going over to an interaction picture the explicit time
dependence of $H_c$ in Eq.
(\ref{bedingter Hamiltonian}) vanishes.
We introduce the operator
\begin{equation}
\label{Generator}
  M\equiv\frac{{\rm i}}{\hbar}H_c=
  \left(
    \begin{array}{cccc}
      0 & {\rm i}\frac{\Omega_L}{2}
      & 0 & -{\rm i}\frac{\Omega_L}{\sqrt{2}}\\
      {\rm i}\frac{\Omega_L}{2} 
      & \frac{\gamma}{2}-{\rm i}\Delta_{2}
      & {\rm i}\Omega & 0\\
      0 & {\rm i}\Omega & -{\rm i}\Delta_{3}
      & -{\rm i}\sqrt{2}\Omega\\
      -{\rm i}\frac{\Omega_L}{\sqrt{2}}
      & 0 & -{\rm i}\sqrt{2}\Omega 
      & \frac{\gamma}{2}-{\rm i}\Delta_{4}
    \end{array}
  \right)
\end{equation}
in matrix form with respect to the atomic basis 
$|1\rangle$, $|2\rangle$, $|3\rangle$, $|4\rangle$,
where $\gamma$ is the Einstein coefficient of the 
Lyman-$\alpha$ transition, $\Omega_L\equiv\frac{e}{\hbar}
F_L\langle2|z|1\rangle$ is the real Rabi frequency of the laser
with respect to the $2p_{\frac{1}{2}}\rightarrow 1s_{\frac{1}{2}}$
transition, $\Omega\equiv\frac{e}{\hbar}F\langle2|z|3\rangle$
is the analogous real constant of the constant electric field
and $\Delta_i\equiv\omega_L-\omega_{i1}$ 
is the detuning of the laser with respect to the 
state $|i\rangle$.
Then we finally have $P_0(t)=\|{\rm e}^{-Mt}|1\rangle\|^2$
\cite{Zerfall de 2s Niveaus}.

\section{The emission behavior}
We assume the laser to be tuned near the 
$2p_{\frac{1}{2}}\rightarrow 1s_{\frac{1}{2}}$
transition resonance, such that $|\Delta_2|\leq\gamma$.
In this case $\hbar |\Delta_3|$ is essentially
given by the Lamb shift between the $2s_{\frac{1}{2}}$
and $2p_{\frac{1}{2}}$ energy level, and $|\Delta_4|$
is a fine structure frequency, which leads to 
$|\Delta_3|\ll|\Delta_4|$. We also assume
the electric fields to be chosen to satisfy the relation
$|\Omega|\ll |\Omega_L|<|\Delta_3|$.
We approximate the function $P_0(t)$
by means of a perturbative approach based on the book of Kato
\cite{Kato}.
Since additionally the $2p_{\frac{3}{2}}$ 
level couples weakly to the laser,
we have two perturbation parameters $\Omega/\Delta_{3}$
and $\Omega_L/\Delta_4$.
Putting $\Omega=0$ in Eq.
(\ref{Generator}) we denote the resulting
operator by $M^{(0)}$,
and we define $M^{(1)}\equiv M-M^{(0)}$.
There exists a decomposition of the form
\begin{equation}
  M^{(0)}=\Lambda_1^{(0)}P_1^{(0)}+\lambda_2^{(0)}P_2^{(0)}
  +\lambda_3^{(0)}P_3^{(0)},
\end{equation}
where $\lambda_2^{(0)}\equiv -{\rm i}\Delta_3$
is one of the eigenvalues, another one is given by 
\begin{equation}
  \label{dritter Eigenwert}
  \lambda_3^{(0)}\approx
  \frac{\gamma}{2}-{\rm i}\Delta_4
\end{equation}
in first order in  $\Omega_L/\Delta_4$
and $P_i^{(0)}$, $i=2,3$ are the respective eigenprojectors,
and we define $P_1^{(0)}\equiv 1-P_2^{(0)}-P_3^{(0)}$.
The operator $\Lambda_1^{(0)}$ is chosen to commute with 
$P_1^{(0)}$. We obtain for  
$\Lambda_1^{(0)}$ in first order in $\Omega_L/\Delta_4$
the result 
\begin{equation}
  \label{zwei Niveau System}
  -{\rm i}\hbar\Lambda_1^{(0)}\approx
  \hbar\frac{\Omega_L}{2}
  \left(|2\rangle\langle 1|+|1\rangle\langle 2|\right)
  -{\rm i}\hbar\left(\frac{\gamma}{2}-{\rm i}\Delta_2\right)
  |2\rangle\langle 2|.
\end{equation}
Note that the right hand side 
of Eq. (\ref{zwei Niveau System}) is the conditional
Hamiltonian of a two-level atom
\cite{Quantensprungmethode,Hegerfeldt reset}.
In an analogous manner we can decompose the operator
$M$ in the form
\begin{equation}
  M=\Lambda_1 P_1 +\lambda_2 P_2+\lambda_3 P_3
\end{equation}
with each $P_i$ corresponding to $P_i^{(0)}$,
and we have 
\begin{equation}
  \label{Zerlegung des Exponentialoperators}
  {\rm e}^{-M t}|1\rangle={\rm e}^{-\Lambda_1 t}P_1|1\rangle 
  +{\rm e}^{-\lambda_2 t}P_2|1\rangle
  +{\rm e}^{-\lambda_3 t}P_3|1\rangle.
\end{equation}
The main idea of our perturbative approach is to approximate 
$\Lambda_1$, $\lambda_2$, $\lambda_3$ and the respective
projectors separately with the aid of Ref.
\cite{Kato}.
First of all we are interested in the behavior of
$P_0(t)$ assuming $t\gg \gamma^{-1}$.
By using Eq. (\ref{dritter Eigenwert})
and Eq. (\ref{zwei Niveau System})
one can verify that the first and the third term
in Eq. (\ref{Zerlegung des Exponentialoperators})
decay exponentially on the time scale $\gamma^{-1}$
while in first order in $\Omega/\Delta_3$
the real part of $\lambda_2$ vanishes.
Because of this we only have to approximate ${\rm Re}(\lambda_2)$
and $\|P_2 |1\rangle\|^2$.
In first order in $\Omega/\Delta_3$
we have
\begin{equation}
  P_2 |1\rangle \approx -P_2^{(0)} M^{(1)} 
  \left(
    M^{(0)}-\lambda_2^{(0)}
    \left(
      1-P_2^{(0)}
    \right)
  \right)
  ^{-1}|1\rangle .
\end{equation}
With the definition of the complex number
\begin{eqnarray}
  \nonumber
  &&\alpha \equiv
  1-\frac{\Delta_3}{\Delta_4}-
  \frac{\Omega_L^2}{4\Delta_3^2}
  -\frac{\Delta_2}{\Delta_3}+\frac{3\Omega_L^2}{4\Delta_3\Delta_4}
  +\frac{\gamma^2}{4\Delta_3\Delta_4}+\frac{\Delta_2}{\Delta_4}\\
  &&-\frac{\Omega_L^2\Delta_2}{2\Delta_3^2\Delta_4}
  -{\rm i}
  \left(
    \frac{\gamma}{2\Delta_3}-\frac{\gamma}{\Delta_4}
    +\frac{3\Omega_L^2\gamma}{8\Delta_3^2\Delta_4}
    +\frac{\gamma\Delta_2}{2\Delta_3\Delta_4}
  \right)
\end{eqnarray}
we obtain
\begin{equation}
  \label{Langzeitverhalten von P0}
  P_0(t)\approx
  {\rm e}^{-2{\rm Re}(\lambda_{2})t}
  \frac{\Omega^{2}\Omega_L^{2}}
  {4\Delta_{3}^{4}}
  \frac{
    \left(
      1-3\frac{\Delta_3}{\Delta_4}+2\frac{\Delta_2}{\Delta_4}
    \right)^2
    +\frac{9\gamma^2}{4\Delta_4^2}}
  {\left|\alpha\right|^2}
\end{equation}
assuming $t\gg \gamma^{-1}$, up to small relative 
deviations of the order $\Omega/\Omega_L$.
In second order in $\Omega/\Delta_3$
one finds
\begin{equation}
  \lambda_2\approx\lambda_2^{(0)}-
  \langle 3|M^{(1)}
  \left(
    M^{(0)}-\lambda_2^{(0)}
    \left(
      1-P_2^{(0)}
    \right)
  \right)^{-1}
  M^{(1)}|3\rangle
\end{equation}
and this leads to
\begin{equation}
  \label{kleiner Eigenwert}
  {\rm Re}\lambda_2\approx\frac{\Omega^2\gamma}{2\Delta_3^2}
  \frac{ 
    1-2\frac{\Delta_3}{\Delta_4}
    +3\frac{\Delta_3^2}{\Delta_4^2}
    -4\frac{\Delta_2\Delta_3}{\Delta_4^2}
    +\frac{3\gamma^2}{4\Delta_4^2}
    +2\frac{\Delta_2^2}{\Delta_4^2}}
  {\left|\alpha\right|^2}.
\end{equation}
It is comparatively easy to describe the behavior of 
$P_0(t)$ on the other time scale $t\not\gg\gamma^{-1}$.
In zeroth order in $\Omega_L/\Delta_4$,
$\Omega/\Delta_3$ we have
\begin{equation}
  \label{Kurzzeitverhalten von P0}
  P_0(t)\approx
  \left\|
    {\rm e}^{-\Lambda_1^{(0)}t}
    |1\rangle
  \right\|^2,
\end{equation}
where $\Lambda_1^{(0)}$ can be approximated by 
Eq. (\ref{zwei Niveau System}).

We introduce a time $T_0$ such that 
$\gamma^{-1}\ll T_0\ll(2{\rm Re}\lambda_2)^{-1}$.
Then for $t\ll T_0$
the function $P_0(t)$ is governed by the behavior of
a two-level atom with a strong transition, while 
in a large time interval around $T_0$ it is very small,
though not vanishingly small, and slowly 
varying.
An interruption of the atomic fluorescence longer
than $T_0$ is called a dark period. The above results concerning 
$P_0(t)$ guarantee the occurrence of light and 
dark periods in the resonance fluorescence of the
atom (see for example
\cite{Emissionsprozess: Cohen Tannoudji}).
Following Refs. 
\cite{Quantensprungmethode,Emissionsprozess: Cohen Tannoudji}
we can calculate the mean durations $T_L$, $T_D$
of the light and dark periods and the probability
$p$ for the occurrence of a dark period.
One finds $p=P_0(T_0)$ and 
$T_D=(2{\rm Re}\lambda_2)^{-1}$,
which is given by Eq. 
(\ref{Langzeitverhalten von P0}), Eq. (\ref{kleiner Eigenwert})
respectively.
The value of $T_L$ can be obtained from
$T_L=\tau_L/p$,
where $\tau_L$ is the mean time between two
photons in a light period.
This is intuitively obvious, since $p^{-1}$
is the mean number of photons in a light period.
We have
\begin{equation} 
  \tau_L=-\int_0^{T_0}t\frac{\dot{P}_0(t)}{1-p}{\rm d}t
  \approx\frac{1}{\gamma}
  \frac{\gamma^2+2\Omega_L^2+4\Delta_2^2}{\Omega_L^2}
\end{equation}
by using Eq.
(\ref{Kurzzeitverhalten von P0}).
Thus one finds
\begin{equation}
  \label{mittlere Dunkelphasenlaenge}
  T_D=\frac{\Delta_3^2|\alpha|^2}
  {\Omega^2\gamma
  \left(
    1-2\frac{\Delta_3}{\Delta_4}
    +3\frac{\Delta_3^2}{\Delta_4^2}
    -4\frac{\Delta_2\Delta_3}{\Delta_4^2}
    +\frac{3\gamma^2}{4\Delta_4^2}
    +2\frac{\Delta_2^2}{\Delta_4^2}
  \right)}
\end{equation}
and
\begin{equation}
  T_L=\frac{4\Delta_{3}^{4}\left|\alpha\right|^2
    \left(
      \gamma^2+2\Omega_L^2+4\Delta_2^2
    \right)}
  {\gamma\Omega_L^4\Omega^2
    \left(
      \left(
        1-3\frac{\Delta_3}{\Delta_4}+2\frac{\Delta_2}{\Delta_4}
      \right)^2
      +\frac{9\gamma^2}{4\Delta_4^2}
    \right)}.
\end{equation}
All the above mean values can be obtained from a single
trajectory of the photon-counting process.

As a typical example of the occurrence of macroscopic
dark periods in the emission process of $^4 {\rm He}^+$
we discuss $\Delta_2=0$,
$F=3.6\times 10^3 \hspace{3pt} \frac{{\rm V}}{{\rm m}}$,
$F_L=2.9\times 10^6 \frac{{\rm V}}{{\rm m}}$,
which means $\Omega=0.025\gamma$, $\Omega_L=5\gamma$.
One finds 
\begin{equation}
  T_D=1.1\times 10^{-5}\hspace{3pt} {\rm s},\quad
  T_L=4\times 10^{-4}\hspace{3pt} {\rm s},
\end{equation}
and for the mean number of photons in a light period
we obtain $p^{-1}=2\times 10^6$.
Since the Lyman-$\alpha$ transition is remarkably strong
with a lifetime of about $0.1\hspace{3pt}{\rm ns}$,
one has a high fluorescence intensity in a light 
period and a different time scale in comparison
with the Dehmelt systems in Ref. 
\cite{Dehmelt System Experimente}.

Under consideration of our premises with respect to the 
electric fields we know that $T_L/T_D$ is 
almost independent of $\Omega$. On the other hand
$T_D$ only depends weakly on $\Omega_L$, which is
intuitively obvious. As a conclusion, one can regulate 
the emission process with the aid of the electric fields.

\section{Discussion}
From the calculation above we have seen that 
macroscopic dark periods occur in hydrogen-like
systems like $^4{\rm He}^+$ provided the
external electric fields are suitably chosen.
One might wonder, however, in which way one 
can reach a dark period.
As the quantum mechanical calculation shows
and the intuitive explanation in the introduction 
suggests, in order to reach a macroscopic
dark period the system must be mostly in the 
$2s$ state.

One might be tempted to argue in a 
simplified way as follows.
One could assume that the coherent evolution
of the atom is started by the 
absorption of a $1s\rightarrow 2p$ photon,
and terminated by spontaneous emission 
into this channel.
In order to evolve to an extended dark period
spontaneous emission must not occur for
many lifetimes.
If we take the effective Rabi
frequency $\Omega=1/40\gamma$ 
of the Stark field
one would estimate
that the probability to obtain an even mixing of 
$2p$ and $2s$ is smaller than $\exp(-20)$, or
$2\times 10^{-9}$, and 
at this point the atom is not dark at all.
From the $0.1\hspace{3pt}{\rm ns}$ 
lifetime of the $2p$ state of $^4{\rm He}^+$
one estimates a lower limit of 
$T_L>0.1\hspace{3pt}{\rm s}$
for the mean time of a light period in the
emission process.
This result much exceeds the previously 
calculated value of 
$T_L=4\times 10^{-4}\hspace{3pt}{\rm s}$
from the quantum mechanical description,
and macroscopic dark periods should be 
very seldom.  

At this point we have to remember that it is the
relative weight of the $2s$ state in the
emission-free subensemble that counts rather than
the absolute population.
There are {\em two} mechanisms that make the $2s$ state
become rapidly predominant in the emission-free
subensemble as follows.
The relative weight of the $2p$ and the $1s$
state in the emission-free subensemble
decreases rapidly on the time scale $\gamma^{-1}$,
because those atoms with a spontaneous
emission from the strong Lyman-$\alpha$ transition
leave the emission-free subensemble
and do not contribute.
On the other hand the $2s$ state is metastable
and weakly coupled to the $2p$ state.
Therefore if the atom is once in the $2s$ state 
it stays with a high probability, and it
remains in the emission-free 
subensemble for a long time.
As a conclusion we obtain the possibly astonishing 
result that the $2s$ state becomes
predominant fairly quickly in the emission-free subensemble
although the absolute population of this
metastable state is very small.
An estimation of the population dynamics  
in the emission-free subensemble  
from a very simple rate equation model
is given in the appendix.
 
For the mechanism of quantum jumps in hydrogen-like 
systems there exists a close relation to the 
proposal of macroscopic dark periods without a
metastable state by Hegerfeldt and Plenio
\cite{Hegerfeld Plenio:
  Dunkelphasen ohne metastabiles Niveau}
as follows.
If we neglect the existence of the weakly coupled
$2p_{\frac{3}{2}}$ level, and if we consider the 
conditional Hamiltonian, which is given in matrix form 
with respect to the atomic 
orthonormal basis $|1\rangle$, 
$\frac{1}{\sqrt{2}}\left(|2\rangle +
|3\rangle\right)$,
$\frac{1}{\sqrt{2}}\left(|2\rangle -
|3\rangle\right)$
by 
\begin{equation}
  \label{Dressed states Formulierung}
  H_c=\hbar
  \left(
    \begin{array}{ccc}
      0 & \frac{\Omega_L}{2\sqrt{2}}
      & \frac{\Omega_L}{2\sqrt{2}}\\
      \frac{\Omega_L}{2\sqrt{2}}
      & \Omega-{\rm i}\frac{\gamma}{4}
      -\frac{\Delta_{2}+\Delta_{3}}{2}
      &-{\rm i}\frac{\gamma}{4}
      +\frac{\Delta_{3}-\Delta_{2}}{2}\\
      \frac{\Omega_L}{2\sqrt{2}}
      &-{\rm i}\frac{\gamma}{4}
      +\frac{\Delta_{3}-\Delta_{2}}{2}
      & -\Omega-{\rm i}\frac{\gamma}{4}
      -\frac{\Delta_{2}+\Delta_{3}}{2}
    \end{array}
  \right),
\end{equation}
then Eq. (\ref{Dressed states Formulierung})
corresponds directly to Eq. (5)
of \cite{Hegerfeld Plenio:
  Dunkelphasen ohne metastabiles Niveau}, except that the 
off-diagonal frequency shift terms 
$\left(\Delta_{3}-\Delta_{2}\right)/2$
(half of the negative Lamb shift frequency) 
are absent
\cite{Altenmueller}.
As a conclusion, the physical system of 
Hegerfeldt and Plenio behaves like a hydrogen-like 
system without quantum electrodynamical corrections
of the atomic spectrum.

In the case of our realistic hydrogen-like system
the photon-counting process is governed 
by the detuning $\Delta_3$, 
where $\hbar |\Delta_3|$ is essentially given 
by the Lamb shift between the $2s_{\frac{1}{2}}$
and the $2p_{\frac{1}{2}}$ energy level.
We assume a possibly realizable measurement
of the mean duration of the dark periods,
as it was done in the experiments 
\cite{Dehmelt System Experimente,alternatives Experiment von
  Hulet und Wineland} for Dehmelt systems.
In this case we can calculate the detuning
$\Delta_3$ with the aid of the approximation Eq.
(\ref{mittlere Dunkelphasenlaenge}) to 
high accuracy
by solving Eq. 
(\ref{mittlere Dunkelphasenlaenge})
for $\Delta_3$ which leads to 
a polynomial equation of sixth degree.
If the other parameters in Eq.
(\ref{mittlere Dunkelphasenlaenge})
are known, this provides a detection of the Lamb shift
by using the proposed photon-counting process.
\section*{Acknowledgments}
I am grateful to G.~C.~Hegerfeldt,
D.~G.~Sondermann, and A.~W.~Vogt
for helpful discussions.

\section*{Appendix}
In this appendix we show that,
in the emission-free subensemble,
the relative weight 
of the $2s$ state becomes
rapidly predominant on the time scale of the inverse 
Lyman-$\alpha$ Einstein coefficient $\gamma^{-1}$.
This can be seen by a simple rate equation model 
for the emission-free subensemble as follows.
For convenience we neglect the existence
of the weakly coupled $2p_{\frac{3}{2}}$ level.
By $P_i(t)$ $(i=1,2,3)$ we denote the 
probability that no photon 
has been detected until $t$ {\em and} 
that the atom is in the state 
$|i\rangle$ at time $t$. 
We note that $t>0$ implies $\sum_{i=1}^3 P_i(t)<1$.
By $R_B$, $R_R$ we denote the transition 
rates due to stimulated 
emission of the blue transition $2p\rightarrow 1s$
and the red transition 
$2s\rightarrow 2p$ respectively.
For this subensemble one has the rate 
equations
\begin{eqnarray}
  \label{Ratengleichungen}
  &&\dot{P}_1=-R_B P_1+R_B P_2,\\
  &&\dot{P}_2=R_B P_1-(\gamma+R_B+R_R) P_2+R_R P_3,\\
  &&\dot{P}_3=R_R P_2-R_R P_3.
\end{eqnarray}
The only difference to the usual rate equations
\cite{Loudon Lehrbuch} 
is that those atoms with a 
spontaneous emission from the 
blue transition leave the 
emission-free subensemble.
Therefore the term $\gamma P_2$ in Eq.
(\ref{Ratengleichungen}) is absent.
This leads to a decay of $\sum_i P_i(t)$ in time.
Assuming $R_R\ll R_B,\gamma$ 
and starting with the ground state
one easily obtains the perturbative expressions
\begin{eqnarray}
  &&P_1(t)=
  {\rm e}^{-\mu_2 t}\frac{\mu_1-R_B}{\mu_1-\mu_2}
  -{\rm e}^{-\mu_1 t}
  \frac{\mu_2-R_B}{\mu_1-\mu_2},\\
  \nonumber
  &&P_2(t)=\frac{R_B}{\mu_1-\mu_2}
  \left(
    {\rm e}^{-\mu_2 t}-{\rm e}^{-\mu_1 t}
  \right),\\
  \nonumber
  &&P_3(t)=\frac{R_R R_B}{\mu_1-\mu_2}
  \left(
    \frac{{\rm e}^{-\mu_1 t}}{\mu_1-\mu_3}
    -\frac{{\rm e}^{-\mu_2 t}}{\mu_2-\mu_3}
  \right)\\
  \nonumber
  &&+{\rm e}^{-\mu_3 t}
  \left(
    \frac{R_R R_B}{(\mu_2-\mu_3)(\mu_1-\mu_2)}-
    \frac{R_R R_B}{(\mu_1-\mu_3)(\mu_1-\mu_2)}
  \right),
\end{eqnarray}
where the real numbers $\mu_i$ are defined
by 
\begin{eqnarray}
  \nonumber
  &&\mu_{1/2}\equiv\frac{1}{2}
  \left(
    (\gamma+R_R+2R_B)\pm
    \sqrt{(\gamma+R_R)^2+4 R_B^2}
  \right),\\
  &&\mu_3\equiv R_R.
\end{eqnarray}
We note that $\mu_1 > \mu_2\gg \mu_3$, and we
see that the population $P_3(t)$ 
of the $2s$ state $|3\rangle$
in the emission-free  
subensemble increases {\em rapidly} from
$P_3(0)=0$
on the time scale 
$\mu_{1/2}^{-1}\sim R_B^{-1}\sim \gamma^{-1}$ 
and then 
remains on a low level for 
a long time of the order
$\mu_3^{-1}\sim R_R^{-1}$.
On the other hand the population $P_1(t)$,
$P_2(t)$
of the $1s$, $2p$ state respectively
decreases {\em rapidly} on the time 
scale $R_B^{-1}\sim \gamma^{-1}$
so that $P_3(t)>P_2(t),P_3(t)$ 
is reached fairly quickly.
This behavior can also be seen in
Fig.\ \ref{fig2}.
Because of the normalization 
($\sum_i P_i(t)=$ weight of the
emission-free subensemble) 
the conditional probabilities are
$P_j(t)/\sum_i P_i(t)$
($j=1,2,3$). 
A similar behavior is obtained 
in the quantum mechanical 
calculation of the paper, but a quantitative
agreement is not easily available from this simple
rate equation model.

\begin{figure}
  \caption{Relevant energy levels of $^4 {\rm He}^+$.
    The fine structure frequency between the $2p_{\frac{3}{2}}$
    and the $2p_{\frac{1}{2}}$ energy level is given by 
    $1.75\times  10^{11}\hspace{3pt}  {\rm Hz}$, 
    while the Lamb shift frequency
    between the $2s_{\frac{1}{2}}$ and the $2p_{\frac{1}{2}}$
    energy levels is $1.4\times  10^{10}\hspace{3pt} {\rm Hz}$.
    In addition $\gamma =10^{10}\hspace{3pt} {\rm s}^{-1}$
    is the Einstein coefficient of the Lyman-$\alpha$
    transition.
    Note that the above Lamb shift splitting is 
    appreciably larger than $\gamma$.
    }
  \label{fig1}
\end{figure}
\begin{figure}
  \caption{Estimation of the expected population dynamics
    by means of the simplified rate equation 
    model in the case of the parameters $R_B=5 \gamma$,
    $R_R=0.05\gamma$.
    The dashed line, fat solid line, thin solid line
    indicate the population $P_1(t)$, $P_2(t)$, $P_3(t)$
    respectively.
    The time axis is given in natural units of the 
    inverse Lyman-$\alpha$ Einstein coefficient.}
  \label{fig2}
\end{figure}

\begin{thebibliography}{2}
\bibitem
  {Andrae: Stark Beats Artikel}
  H. J. Andr\"a,
  Phys. Rev. A {\bf 2},
  2200 (1971);
  H. J. Andr\"a, {\it Fast-Beam
    (Beam-Foil) Spectroscopy}
  in {\it Progress in Atomic Spectroscopy},
  Part B, edited by W. Hanle
  and H. Kleinpoppen, Plenum (1979), p. 893.
\bibitem
  {Dehmelt Quantenspruenge}
  H. G. Dehmelt, Bull. Am. Phys.
  {\bf 20}, 60 (1975).
\bibitem
  {Dehmelt System Experimente}
  W. Nagourney, J. Sandberg, and H. Dehmelt,
  Phys. Rev. Lett. {\bf 56},
  2797 (1986);
  J. C. Bergquist, R. G. Hulet,
  W. M. Itano, and D. J. Wineland,
  Phys. Rev. Lett. {\bf 57},
  1699 (1986);
  Th. Sauter, W. Neuhauser, R. Blatt, and 
  P. E. Toschek,
  Phys. Rev. Lett. {\bf 57},
  1696 (1986).
\bibitem
  {QM Beschreibung der Leucht und Dunkelphasen}
  A. Schenzle, R. G. DeVoe, and R. G. Brewer,
  Phys. Rev. A {\bf 33}, 2127 (1986);
  J. Javanainen, Phys. Rev. A {\bf 33},
  2121 (1986);
  D. T. Pegg, R. Loudon, and P. L. Knight,
  Phys. Rev. A {\bf 33}, 4085 (1986);
  G. Nienhuis, Phys. Rev. A {\bf 35},
  4639 (1987);
  P. Zoller, M. Marte, and D. J. Walls,
  Phys. Rev. A {\bf 35}, 198 (1987);
  M. Porrati and S. Putterman,
  Phys. Rev. A {\bf 36}, 925 (1987);
  S. Renaud, J. Dalibard, and C. Cohen-Tannoudji,
  IEEE Journal of Quantum Electronics $\mathbf{24}$,
  1395 (1988);
  D. T. Pegg and P. L. Knight,
  Phys. Rev A {\bf 37}, 4303 (1988).
\bibitem
  {Quantensprungmethode}
  T. S. Wilser, Ph. D. dissertation,
  G\"ottingen (1991);
  G. C. Hegerfeldt, T. S. Wilser,
  {\it Proceedings of the II. 
    International Wigner Symposium},
    Goslar (1991), edited by H. D. Doebner,
  W. Scherer, and F. Schroeck,
  World Scientific, Singapore (1992). 
\bibitem
  {Emissionsprozess: Cohen Tannoudji}
  C. Cohen-Tannoudji and J. Dalibard,
  Europhys. Lett.
  {\bf 1}, 441 (1986). 
\bibitem
  {alternatives Experiment von
    Hulet und Wineland}
  R. G. Hulet, D. J. Wineland,
  Phys. Rev. A {\bf 36},
  2758 (1987).
\bibitem
  {Hegerfeld Plenio:
    Dunkelphasen ohne metastabiles Niveau}
  G. C. Hegerfeldt, M. B. Plenio,
  Phys. Rev. A {\bf 46},
  373 (1992).
\bibitem
  {Hegerfeldt Plenio:
    Dunkelphasen in
    Lambda-Systemen}
  G. C. Hegerfeldt, M. B. Plenio,
  Z. Phys. B {\bf 96},
  533 (1995).
\bibitem
  {Cook Kimble: Telegraphenprozess} 
  H. J. Kimble, R. J. Cook, and A. L. Wells,
  Phys. Rev. A {\bf 34}, 3190 (1986)
  have used a Bloch equation approach to the Dehmelt
  system which might be carried over to the present
  case. However, as the authors point out their
  approximations become problematic for a large 
  detuning of the weak transition.
\bibitem
  {Lamb Retherford}
  W. E. Lamb, R. C. Retherford, Phys. Rev.
  {\bf 72}, 241 (1947).
\bibitem
  {Hyperfeinstruktur}
  In the case of $^4 {\rm He}^+$ there is 
  no hyperfinestructure since the nucleus consists of 
  an even number of both neutrons and protons.
\bibitem
  {Bethe Salpeter}
  H. A. Bethe, E. E. Salpeter,
  {\it Quantum Mechanics of One-and
  Two-Electron Systems}, in
  {\it Encyclopedia of Physics}, Volume 35,
  Atoms I, edited by
  S. Fl\"ugge, Springer (1957).
\bibitem
  {Loudon Lehrbuch}
  R. Loudon, {\it The quantum theory of light},
  Clarendon (1983).
\bibitem
  {Davies Srinivas}
  M. D. Srinivas and E. B. Davies,
  Opt. Acta {\bf 28},
  981 (1981).
\bibitem
  {Gardiner: Quantum noise}
  C. W. Gardiner,
  {\it Quantum Noise},
  Springer (1991).
\bibitem
  {Hegerfeldt reset}
  G. C. Hegerfeldt, 
  Phys. Rev. A
  {\bf 47},
  449 (1993).
\bibitem
  {MCWA}
  J. Dalibard, Y. Castin and
  K. M{\o}lmer, Phys. Rev. Lett.
  $\mathbf{68}$, 580 (1992);
  K. M{\o}lmer, Y. Castin
  and J. Dalibard, J. Opt. Soc. Am. 
  B {\bf 10}, 524 (1993).
\bibitem
  {Carmichael Quantum Trajectories}
  H. J. Carmichael,
  {\it An Open System Approach to
    Quantum Optics},
  Lecture Notes in Physics,
  Springer (1993).
\bibitem
  {Erklaerung zum Ruecksetzoperator}
  If we assume a single ground state this 
  would also be the state right after the detection of
  a photon, and in this case it clearly does not depend on
  the state before.
\bibitem
  {Darstellungstheorie}
  Here we use the invariant measure
  (see for example M. A. Naimark, A. I. $\check{\rm S}$tern,
  {\it Theory of Group Representations},
  Springer (1982)) of the
  compact group $O(2)$.
  It is also possible to find the same results 
  by a purely algebraic reasoning. In this paper we 
  choose the simplest calculation.
\bibitem
  {Verallgemeinerung}
  By a similar reasoning one can generalize
  the approach of dimensional reduction
  to the case of a unitary representation of an
  arbitrary symmetry group.
\bibitem
  {Lueders Lebensdauer}
  G. L\"uders, Z. Naturforschg. 
  {\bf 5a}, 608-611 (1950).
\bibitem
  {Zerfall de 2s Niveaus}
  For convenience we neglect the spontaneous decay of
  the $2s$ state. 
  For the example of $^4 {\rm He}^+$
  we have a lifetime of about $2\hspace{3pt}{\rm ms}$.
\bibitem 
  {Kato}
  T. Kato, {\it Perturbation theory for linear operators},
  Springer (1966), p. 62.
\bibitem
  {Altenmueller}
  Cf. also
  T. P. Altenm\"uller,
  Z. Phys. D
  {\bf 34}, 157 (1995).
\end{thebibliography}
\end{document}